# Rotating vortex clusters in media with inhomogeneous defocusing nonlinearity

**Yaroslav V. Kartashov,**[1,2,*] **Boris A. Malomed,**[3,4] **Victor A. Vysloukh,**[5] **Milivoj R. Belić,**[6] **and Lluis Torner**[1,7]

[1]*ICFO-Institut de Ciencies Fotoniques, The Barcelona Institute of Science and Technology, 08860 Castelldefels (Barcelona), Spain*
[2]*Institute of Spectroscopy, Russian Academy of Sciences, Troitsk, Moscow Region, 142190, Russia*
[3]*Department of Physical Electronics, School of Electrical Engineering, Faculty of Engineering, Tel Aviv University, 69978 Tel Aviv, Israel*
[4]*Laboratory of Nonlinear-Optical Informatics, ITMO University, St. Petersburg 197101, Russia*
[5]*Universidad de las Americas Puebla, Santa Catarina Martir, 72820, Puebla, Mexico*
[6]*Science Program, Texas A&M University at Qatar, P.O. Box 23874 Doha, Qatar*
[7]*Universitat Politècnica de Catalunya, 08034, Barcelona, Spain*

**We show that media with inhomogeneous defocusing cubic nonlinearity growing toward the periphery can support a variety of stable vortex clusters nested in a common localized envelope. Nonrotating symmetric clusters are built of an even number of vortices with opposite topological charges, located at equal distances from the origin. Rotation makes the clusters strongly asymmetric, as the centrifugal force shifts some vortices to the periphery, while others approach the origin, depending on the topological charge. We obtain such asymmetric clusters as stationary states in the rotating coordinate frame, identify their existence domains, and show that the rotation may stabilize some of them.**

The existence of two- and three-dimensional (2D) and (3D) self-sustained topological states is a topic of continuously renewed interest in nonlinear optics [1,2], Bose-Einstein condensates [3,4], field theory [5], and other fields. Unlike their 1D counterparts, most 2D and 3D soliton states are subject to strong instabilities [6], vortex solitons being particularly prone to azimuthal instabilities. Finding physically relevant settings which may stabilize them has drawn much interest. Among other conceptual approaches, one that eliminates collapse and also helps to suppress splitting instabilities is based on the use of a defocusing cubic nonlinearity whose strength grows to periphery at a rate faster than $r^D$, where $r$ is the radial coordinate and $D$ the dimension of space [7]. Such a model predicted the existence of stable 3D complex states, such as hopfions [8]. Various forms of stationary 2D soliton patterns in inhomogeneous nonlinearity landscapes were studied too [9-12]. In addition to single vortices, they include *nonrotating* clusters built of vortices and antivortices nested in a confined wave field, such as vortex-antivortex dipoles and quadrupoles [11,12]. Vortex clusters are objects of special interest but, due to their structural complexity, they are often unstable, or feature complex alternating stability-instability domains [11]. The rich dynamics of individual vortices [13,14] and vortex clusters [15-18] has been also studied in Bose-Einstein condensates held in parabolic traps. In symmetric traps, only vortex dipoles were found to be stable [15]. The stabilization of line clusters by the asymmetry of the trap was reported in [16], *azimuthons* containing vortices were constructed in [17], and their possible excitation was discussed in [18].

However, steadily rotating vortex clusters have never been obtained in an accurate form in a medium with inhomogeneous defocusing nonlinearity. In this Letter we study such states and elucidate their stability as well as the impact of rotation of their properties. We show that rotation is beneficial, as it may lead to stabilization of spinning patterns. In particular, we found that vortex quadrupoles that become strongly asymmetric due to rotation may be stabilized when the rotation frequency exceeds a certain minimum value. Rotation also causes a specific deformation of the dipoles, and drives a transition between a vortex dipole and a single axially-symmetric vortex placed at the center.

The evolution of light beams in media with inhomogeneous defocusing cubic nonlinearities is described by the nonlinear Schrödinger equation (NLSE) for the scaled field amplitude $q$ [7,19]:

$$i\frac{\partial q}{\partial z} = -\frac{1}{2}\left(\frac{\partial^2 q}{\partial x^2} + \frac{\partial^2 q}{\partial y^2}\right) + \sigma(x,y)|q|^2 q, \qquad (1)$$

where the transverse coordinates $x, y$ are normalized to the input beam width, while the propagation distance $z$ is normalized to the diffraction length. We assume an axially symmetric defocusing ($\sigma > 0$) nonlinearity with the strength growing toward the periphery as $\sigma(r) = \exp(\alpha r^2)$, where $r^2 = x^2 + y^2$, and $\alpha = 0.5$, is set by rescaling. Several methods may potentially be used to create such nonlinearity profiles, including inhomogeneous doping of photorefractive materials [20], selective infiltration of holes in photonic crystal fibers with index-matching liquids [21,22], and utilization of the

Feshbach resonance imposed by nonuniform fields in the case of matter waves [23].

A specific feature of model (1) is its *nonlinearizability* for the tails of localized solutions decaying at $r \to \infty$, due to the growth of the nonlinearity. We look for solutions of the form $q = w \exp(ibz)$, with a complex stationary wave amplitude $w(x,y)$ and real propagation constant $b$. The Thomas-Fermi approximation (TFA), which neglects the diffraction term in Eq. (1), yields $|w|^2 = -b/\sigma(r)$. This expression becomes asymptotically exact at $r \to \infty$, and is valid for solutions of all the types, including vortex clusters [7]. Note that the decay rate of the tails is determined solely by the nonlinearity profile and does not depend on the propagation constant.

Our aim is to find vortex clusters rotating with angular frequency $\omega$, represented by stationary solutions $q = w(x', y') \exp(ibz)$ in the rotating coordinate frame, $x' = x \cos(\omega z) + y \sin(\omega z)$, $y' = y \cos(\omega z) - x \sin(\omega z)$. In this case, Eq. (1) takes the form (we omit primes):

$$i\frac{\partial q}{\partial z} = -\frac{1}{2}\left(\frac{\partial^2 q}{\partial x^2} + \frac{\partial^2 q}{\partial y^2}\right) + i\omega\left(x\frac{\partial q}{\partial y} - y\frac{\partial q}{\partial x}\right) + \sigma(x,y)|q|^2 q, \quad (2)$$

with the Coriolis term $\sim \omega$ ($\omega > 0$ corresponds to counterclockwise rotation). Equation (2) was solved with the help of the Newton algorithm. We seek for clusters that, at $\omega \to 0$, are composed of $2n$ vortices with alternating charges $\pm 1$, symmetrically placed along a ring, and consider the case of a single vortex separately. For each $n$ there exist only one soliton family of this type parameterized by $b$ and $\omega$ that is symmetric at $\omega = 0$ and becomes asymmetric at $\omega > 0$. Examples of such nonrotating and rotating clusters are displayed in Fig. 1. They are characterized by the energy flow (norm), $U = \iint |q|^2 \, dx dy$, which is considered a function of propagation constant $b$ and $\omega$ (the TFA yields $U = -\pi b/\alpha$). Under the inhomogeneous defocusing nonlinearity, solitons exist for $b < 0$ [7].

The simplest rotating state features precession [24,25] (circular motion) of a single vortex, as shown in Fig. 1(a). It should be stressed that, in contrast to all other structures displayed in Fig. 1, this state is not represented by a (numerically) exact stationary soliton. It appears as a dynamical state (which is consistent with the findings of Ref. [13] for parabolic trapping potentials) featuring indefinitely long persistent rotation with minimal amplitude oscillations, whose shape, nevertheless, can be found only with a limited accuracy (a small iteration error cannot be reduced below a certain level). We thus characterize such states by the energy flow $U$, rather than by the propagation constant $b$, as the latter may be associated only with truly stationary solitons. For a fixed $U$, the rotation frequency of the single vortex only slightly varies with the change in the offset of its pivot from the origin, $x_+$, in sharp contrast to the precession of a single vortex in parabolic trapping potentials, where the frequency definitely decreases with the increase of $x_+$ [13]. On the other hand, $\omega$ increases with the growth of the energy flow at fixed $x_+$ [Fig. 2(a)].

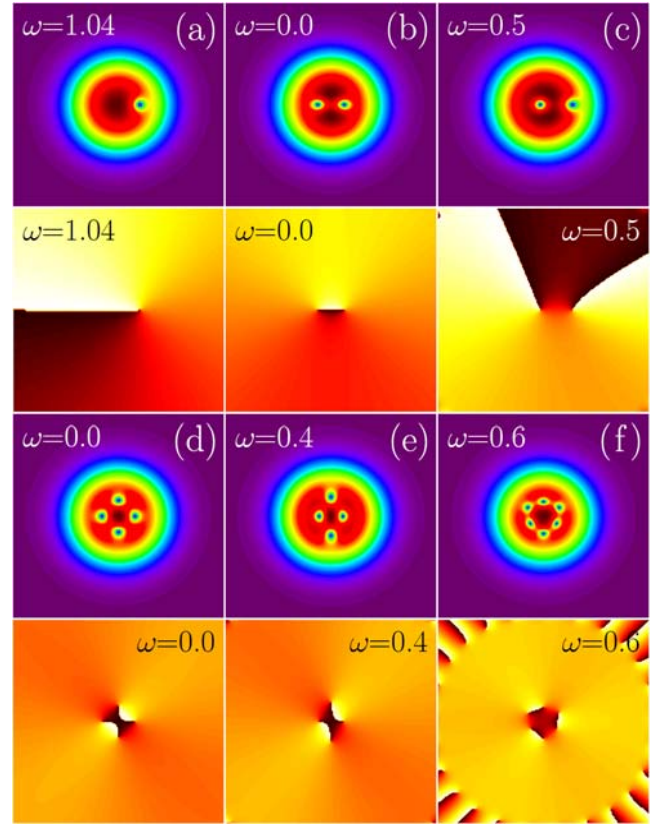

Fig. 1. (Color online) Distributions of the field absolute value (top) and phase (bottom) in vortices and vortex clusters. (a) A dynamical state with a single pivot at $\omega = 1.04$, $U = 58.4$; (b) and (c): vortex dipoles with $b = -10$ and $\omega = 0$ or $\omega = 0.5$, respectively; (d) and (e): vortex quadrupoles with $b = -10$ and $\omega = 0$ or $\omega = 0.4$, respectively; (f) a vortex sextupole with $b = -20$, $\omega = 0.6$ (note that the symmetry of the latter structure is reduced from hexagonal to triangular). All the patterns are shown in the domain $x, y \in [-5, +5]$.

Two oppositely charged vortices build a dipole [25], which can be found as a numerically exact stationary solution in the rotating reference frame [Eq. (2)]. It is symmetric in the absence of rotation [Fig. 1(b)]. When set in rotation, it becomes asymmetric [Fig. 1(c)] with a degree of asymmetry that increases with $\omega$. Namely, for $\omega > 0$, the right pivot in Fig. 1(c) moves to the periphery, while the left one gradually shifts to the origin. The picture is inverted for $\omega < 0$. This fact indicates that the centrifugal force in the rotating frame depends on the topological charge of the vortex and results in different displacements of vortices with $m = \pm 1$. Figure 2(b) illustrates the variation of positions $x_+$ and $x_-$ of the right and left pivots upon the increase of $\omega$ for a fixed propagation constant $b$. When $\omega$ approaches the maximal value, one vortex moves to the transverse infinity, where it eventually disappears, while the other falls onto the center, so that the rotating vortex dipole transforms into a radially symmetric vortex soliton with charge $m = -1$, which is also associated with frequency $\omega$. A similar transformation occurs upon the variation of the propagation constant $b$ at fixed $\omega$ [Fig. 2(c)]. The strongest asymmetry of the dipole shape is observed at small values of $|b|$, while for $b \to -\infty$ the separation between the pivots and overall asymmetry of the dipole decrease. Thus, for fixed $\omega$, Fig. 2(c) shows that there exists a certain minimum value $|b|_{\min}$ of $|b|$, which corresponds to $x_+ \to \infty$, $x_- = 0$, with no vortex dipoles existing for $|b| < |b|_{\min}$.

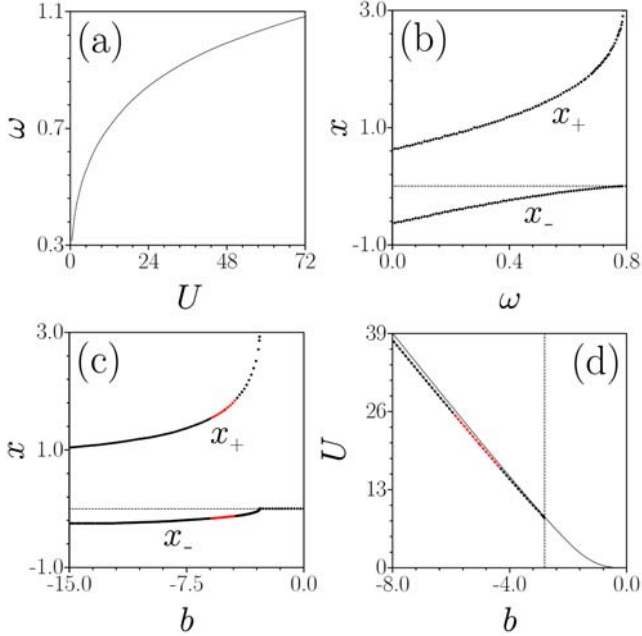

Fig. 2. (Color online) (a) Rotation frequency $\omega$ for the dynamical state with a single vortex, located at $x_+ = 1$, versus energy flow. (b) and (c): The variation of positions of pivots in the vortex dipole with the increase of $\omega$ at $b = -10$ and with the decrease of $b$ at $\omega = 0.4$, respectively. (d) $U(b)$ dependence, illustrating the bifurcation of a vortex dipole (line with circles) from the axially symmetric vortex soliton (straight line) with charge $m = -1$ at $\omega = 0.4$. In (b)-(d) stable and unstable branches are black and red, respectively.

The energy flow of the vortex dipole only slightly and monotonously decreases with the increase of $\omega$. The $U(b)$ dependence is more informative [Fig. 2(d)], since it shows that the vortex dipole *bifurcates*, at $|b| = |b|_{\min}$, from an axially-symmetric vortex soliton with charge $m = -1$. Note that just this vortex falls onto the center with the decrease of $|b|$, as shown in Fig. 2(c), while the vortex with $m = +1$ moves to the periphery, at $\omega > 0$ (for $\omega < 0$ the picture is opposite). We stress that in the rotating frame the dependences $U(b)$ for axially symmetric solitons with opposite charges split. Indeed, Eq. (2) rewritten in polar coordinates $(r, \theta)$ in the rotating frame, reads

$$i\frac{\partial q}{\partial z} - i\omega \frac{\partial q}{\partial \theta} = -\frac{1}{2}\left(\frac{\partial^2 q}{\partial r^2} + \frac{1}{r}\frac{\partial q}{\partial r} + \frac{1}{r^2}\frac{\partial^2 q}{\partial \theta^2}\right) + \sigma(r)|q|^2 q. \quad (3)$$

The substitution $q = w(r)\exp(im\theta + ibz)$ for the vortex soliton yields the term $(m\omega - b)w$ on the left-hand side of Eq. (3), making readily apparent that the corresponding dependences $U(b)$ for the vortex solitons with opposite charges are mutually shifted by $2|m\omega|$ in the horizontal direction (which is the above-mentioned split). For $\omega > 0$, the bifurcation of the vortex dipole occurs only from one of these dependencies, *viz.*, the one associated with $m = -1$.

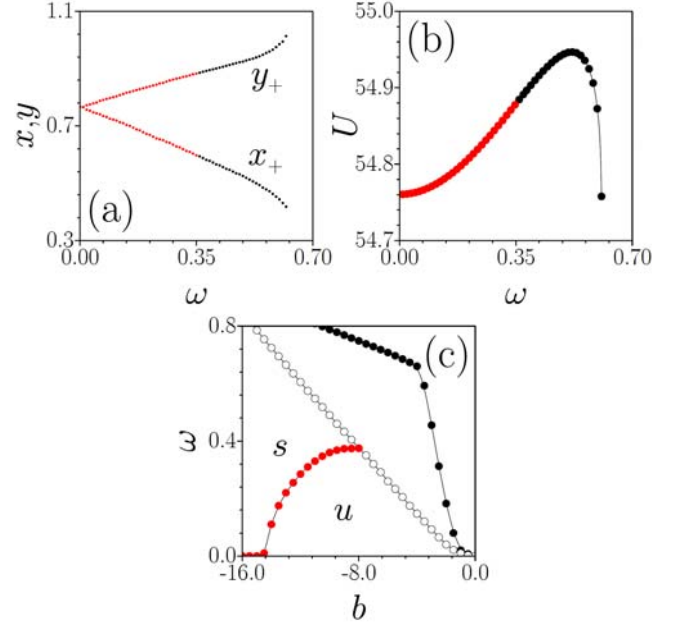

Fig. 3. (Color online) Variation of pivot positions $x_+, y_+$ in a vortex quadrupole (a), and its energy flow (b) versus $\omega$ at $b = -12$. Two other pivots are located at $x_- = -x_+$ and $y_- = -y_+$. Stable and unstable branches are shown in black and red, respectively. (c) Maximal rotation frequency versus $b$ for vortex dipoles and quadrupoles (black and white circles, respectively). Vortex quadrupoles are stable in the domain labelled "s" between the red and white circles, and unstable in the "u" domain.

The next-order cluster, built of three phase dislocations, may be stationary only when all three pivots are set within a common diameter, with the central vortex having the charge opposite to that of the outermost vortices. This state is strongly unstable, so we proceed to quadrupoles composed of four vortices, as those shown in Figs. 1(d) and 1(e). Being symmetric at $\omega = 0$, the quadrupoles become asymmetric (rhombic) at $\omega \neq 0$. For $\omega > 0$, two pivots located on the $x$-axis gradually approach each other, while ones sitting on the $y$-axis gradually escape to periphery, as shown in Fig. 3(a). Although this picture suggests that the quadrupoles might transform into axially-symmetric charge-2 vortices with the increase of $\omega$, this is not the case. Instead, it is observed that a line tangential to the non-monotonous $U(\omega)$ dependence for the quadrupoles becomes vertical at a certain maximal value of $\omega$, and no further shape transformation occurs [Fig. 3(b)]. This is an indication of the existence of another, more complex, soliton family that merges with the quadrupole family from below (we do not show it here). Figure 3(c) shows existence domains for vortex dipoles and quadrupoles in the $(b, \omega)$ plane. Both domains expand with the increase of $|b|$.

Similar existence domains are found for higher-order rotating vortex clusters, including the sextupole shown in Fig. 1(f). While at $\omega = 0$ it contains six vortices uniformly distributed along a ring, at $\omega \neq 0$ the structure features a pronounced triangular shape, as vortices with positive charges move towards the origin, while the ones with negative charges drift to periphery. Clusters with odd numbers of vortices were not found in the form of rings, rearranging themselves upon iterations into more complex stationary patterns that are unstable.

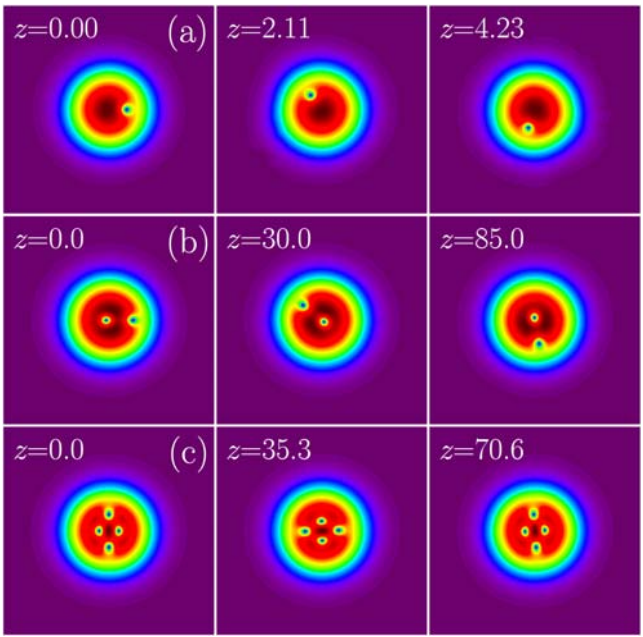

Fig. 4. (Color online) Examples of the evolution of vortex clusters. (a) Dynamical state with one dislocation, $\omega=1.04$, $U=58.4$; (b) Vortex dipole with $\omega=0.5$, $b=-10$; (c) Vortex quadrupole with $\omega=0.4$, $b=-10$. All these states are found to be dynamically stable.

We have tested the stability of vortex clusters by direct numerical propagation, up to $z=10^3$, of the states with weak random input noise added to them. We have thus found narrow instability domains for vortex dipoles [red segments in Figs. 2(c) and (d)]. Such domains are encountered only for small values of $|b|$, and disappear completely when $|b|$ becomes large enough. The instability of the vortex dipoles was found to be always oscillatory. It leads to displacement of the pivots, one of which moves to the periphery and disappears in the regions of vanishing field, while the other one falls to the center, transforming the dipole into the usual isotropic vortex. Vortex quadrupoles are unstable at small values of $|b|$. A stability domain, between the red and white circles in Fig. 3, was found to open at sufficiently large $|b|$, for values of $\omega$ close to the upper border of the existence domain [see also the unstable red and stable black segments in Figs. 3(a)], and it becomes broader with the increase of $|b|$. The opening of such stability domain above a critical value of $\omega$ is an indication of the stabilizing action of the rotation on the vortex clusters. Above a certain value of $|b|$, the vortex quadrupoles become stable for any rotation frequency within their existence domain. We did not obtain truly stable sextupole solutions – we did not explore the whole parameter space – but the decay distance for them was also found to notably increase with the increase of the rotation frequency.

Examples of the stable rotation of a single vortex, and, on the other hand, of vortex dipoles and quadrupoles, corresponding to the numerically exact solutions in the rotating frame, are shown in Fig. 4. The fact that the single orbiting vortex does not correspond to an exact stationary solution causes small-amplitude oscillations developing at a periphery of the mode upon its propagation, which do not vanish even after hundreds of rotation periods. Such oscillations do not appear for dipoles and quadrupoles. Finally, we note that it is interesting to analyze in detail the cluster dynamics in terms of coupled equations of motion for individual vortices [26], which is a subject of a separate work.

Summarizing, we have shown that the medium with inhomogeneous defocusing nonlinearity can support stable rotating vortex clusters. The rotation makes such structures strongly asymmetric, while at the same time stabilizes some families, such as vortex quadrupoles.

We acknowledge support from the Severo Ochoa Excellence program of the Government of Spain, by Fundació Cellex and Fundació Mir-Puig. Work at the Texas A&M University at Qatar is supported by the NPRP 8-028-1-001 project of the Qatar National Research Fund (a member of the Qatar Foundation). The work of B.A.M. is supported, in part, by Grant No. 2015616 from the joint program in physics between NSF and Binational (US-Israel) Science Foundation.